\documentclass[aps,superscriptaddress,twocolumn,nofootinbib]{revtex4}

\makeatletter

\usepackage{natbib}
\usepackage{dcolumn}
\usepackage{graphicx}
\usepackage{amsmath}
\usepackage[hang,small,bf,centerlast]{caption} 
\begin{document}
\title{Cluster analysis for portfolio optimization}

\author{Vincenzo Tola\footnote{Opinions expressed in this paper are exclusively by the authors
and do not necessarily reflect those of Banca d'Italia}}
\affiliation{Dipartimento di Economia, Universit\`a Politecnica delle Marche, Piazza Martelli 8, I-60121 Ancona, Italia}
\affiliation{Banca d'Italia, Rome, Italy}
\author{Fabrizio Lillo}
\affiliation{Dipartimento di Fisica e Tecnologie Relative, Universit\`a di Palermo, viale delle Scienze, I-90128 Palermo, Italia\\
and\\
INFN, Sezione di Catania, Catania, Italy}
\affiliation{Santa Fe Institute, 1399 Hyde Park Road, Santa Fe NM 87501, USA}
\author{Mauro Gallegati}
\affiliation{Dipartimento di Economia, Universit\`a Politecnica delle Marche, Piazza Martelli 8, I-60121 Ancona, Italia}
\author{Rosario N. Mantegna}
\affiliation{Dipartimento di Fisica e Tecnologie Relative, Universit\`a di Palermo, viale delle Scienze, I-90128 Palermo, Italia\\
and\\
INFN, Sezione di Catania, Catania, Italy}

\begin{abstract}
We consider the problem of the statistical uncertainty of the correlation matrix in the optimization of a financial portfolio. We show that the use of clustering algorithms can improve the reliability of the portfolio in terms of  the ratio between predicted and realized risk. Bootstrap analysis indicates that this improvement is obtained in a wide range of the parameters $N$ (number of assets) and $T$ (investment horizon). The predicted and realized risk level and the relative portfolio composition of the selected portfolio for a given value of the portfolio return are also investigated for each considered filtering method.    
\end{abstract}
\maketitle 


\medskip 


\section{Introduction}
The problem of portfolio optimization is one of the most important issue in asset management \cite{Elton95}. Since the seminal work of Markowitz \cite{Markowitz59}, which solved the problem under a certain number of simplifying assumptions (see also Section~\ref{portopt}), many other studies have been devoted to 
consider several aspects of portfolio optimization both from a theoretical and from an applied point of view.
Here we focus our attention on the role of correlation coefficient matrix in portfolio optimization. The estimation of the correlation matrix has unavoidably associated a statistical uncertainty, which is due to the finite length of the asset return time series. Recently, there have been several contributions in the econophysics literature devoted to quantify the degree of statistical uncertainty present in a correlation matrix. The results of these investigations have been obtained by using concepts and tools of random matrix theory (RMT) \cite{metha90}. The RMT quantification of the statistical uncertainty associated with the estimation of the correlation coefficient matrix of a finite multivariate time series has been recently used to device a procedure to filter the information present in the correlation coefficient matrix which is robust with respect to the unavoidable statistical uncertainty (in the econophysics literature it has been used the term of noise dressing) \cite{Galluccio98,Laloux99,Plerou99,Laloux00,Gopikrishnan01,Drozdz01,Rosenow02,Plerou02,Pafka03,Rosenow03,Guhr03,Malevergne04,Sharifi04,Pafka04,Burda04}. The correlation matrices obtained by this filtering procedure has been used in portfolio optimization. Some studies \cite{Laloux00,Rosenow02} have shown that under the assumption of perfect forecasting of future returns and volatilities the distance between the predicted optimal portfolio and the realized one is smaller for the filtered correlation matrix than for the original one at a given level of the portfolio return.

In recent years, other filtering procedures of the correlation coefficient matrix performed using correlation based clustering procedures has also been proposed in the econophysics literature \cite{Mantegna99,Kullmann00,Bonanno00,Giada01,Bonanno01,Maslov01,Bernaschi02,Kullmann02,Onnela02,Mendes03,Micciche03,Maskawa03,Bonanno03,Bonanno04,Dimatteo04,Basalto05,Tumminello05}. These methods also select information of the correlation coefficient matrix which is representative of the entire matrix and it is often less affected by the statistical uncertainty and therefore more stable than the entire matrix during the time evolution of the system.

In this paper we investigate how the portfolio optimization procedure is sensitive to different filtering procedures applied to the correlation coefficient matrix. Specifically we consider filtering procedures based on RMT and on correlation based clustering procedures.  
The paper is organized as follows. In Section~\ref{portopt} we describe briefly the mean variance optimization problem, we define the notation and we summarize the problem of the estimation of the correlation matrix. In Section~\ref{rmtapp} we review the approach recently introduced \cite{Laloux00,Rosenow02}  which makes use of the RMT to improve the portfolio optimization in the presence of estimation errors due to the finiteness of sample data. In Section~\ref{mst} we describe the clustering algorithms used to perform the portfolio optimization. These algorithms are average linkage and single linkage.  In Section~\ref{method} we describe two methods based on these clustering algorithms to build asset portfolios which are robust and reliable. Finally in Section~\ref{conclusions} we summarize our results and indicate future work extending and possibly improving our method. 

\section{Portfolio optimization}\label{portopt}

\subsection{Markowitz's solution}
In this section we briefly discuss the basic aspects of Markowitz portfolio optimization. This is also useful to set the notation and to state the assumption made and the methods used.
Given $N$ risky assets the portfolio composition is determined by the weights $p_i$ ($i=1,...,N$) giving the fraction of wealth invested in asset $i$. 
The weights are normalized as $\sum_{i=1}^{N}p_i=1$. The average return and the variance of the portfolio are
\begin{eqnarray}
r_p\equiv\sum_{i=1}^Np_i m_i \\
\sigma_p^2\equiv\sum_{i=1}^N\sum_{j=1}^Np_ip_j \sigma_{ij},
\end{eqnarray}
where $m_i$ is the mean return of asset $i$ and $\sigma_{ij}$ is the covariance between returns of asset $i$ and $j$. The optimization problem consists in finding the vector ${\bf p}$ which minimizes $\sigma_p$ for a given value of $r_p$. We assume that short selling is allowed, i.e. $p_i$ can assume negative values. As known the solution of this optimization problem has been found by Markowitz \cite{Markowitz59} and it is 
\begin{equation}
{\bf p^*}=\lambda {\bf \Sigma}^{-1}{\bf 1}+\gamma{\bf \Sigma}^{-1}{\bf m}
\label{markowitz}
\end{equation}
where ${\bf \Sigma}$ is the covariance matrix, ${\bf 1}^T=(1,...,1)$ and ${\bf m}$ is the vector of the mean returns of the $N$ assets. The other parameters are
\begin{eqnarray}
\lambda=\frac{C-r_p B}{\Delta}~~~~~~~~~\gamma=\frac{r_pA-B}{\Delta}\nonumber \\
A={\bf 1}^T{\bf \Sigma}^{-1}{\bf 1}~~~~~~B={\bf 1}^T{\bf \Sigma}^{-1}{\bf m}\nonumber\\
C={\bf m}^T{\bf\Sigma}^{-1}{\bf m}~~~~~~\Delta=AC-B^2\nonumber
\end{eqnarray}

\subsection{Curse of dimensionality and adopted method}
The Markowitz's solution to the optimization problem relies upon a series of assumptions that are rarely observed in practice. First of all the asset returns are assumed to be Gaussian variables whereas fat tails in price return distribution are observed. Second the parameters used in the optimization, i.e. the mean values ${\bf m}$ and the covariance matrix ${\bf \Sigma}$, are assumed constant. Finally even if these quantity are really constant in the time horizon relevant for the problem, their statistical estimation over finite time intervals $T$ leads to the problem known as {\it curse of dimensionality}. Since the covariance matrix has $N(N-1)/2\sim N^2/2$ distinct entries whereas the number of records used in the estimation is $NT$, one needs time series of length $T>>N$ in order to have small error on the covariance. But for long $T$ non stationarity becomes more and more important. For these reasons it is important to develop methods able to filter the part of the covariance matrix which is less likely to be affected by statistical uncertainty, and use (when possible) the filtered information to build portfolios.

In this paper we are mainly concerned with the problems in the portfolio optimization due to the estimation of the correlation matrix, i.e the matrix whose entries are the correlation coefficient between returns of different assets. The correlation coefficient is defined as $\rho_{ij}\equiv \sigma_{ij}/\sqrt{\sigma_{ii}\sigma_{jj}}$. Therefore we will use the following procedure \cite{Laloux00,Rosenow02}  to assess the effectiveness of the filtering procedure of the correlation coefficient matrix based on RMT. Given $N$ assets, a portfolio horizon of $T$ trading days and a time $t_0$ when the optimization is supposed to take place, we compute the correlation matrix in the $T$ days preceding  $t_0$ but we compute the mean returns $m_i$ and the volatilities $\sigma_i\equiv \sqrt{\sigma_{ii}}$ in the $T$ days following $t_0$. We use these data to compute the covariance matrix and the predicted optimal portfolio at time $t_0$. We then compare the predicted risk-return curve with the realized risk-return curve obtained by computing   
\begin{equation}   
\hat \sigma_p^2\equiv\sum_{i=1}^N\sum_{j=1}^Np^*_ip^*_j \hat\sigma_{ij},
\label{realizedrisk}
\end{equation}
where $p^*_i$ are the weights obtained in the optimization and $\hat\sigma_{ij}$ is the covariance matrix observed between $t_0$ and $t_0+T$. By using this procedure we are able to decouple the  problem of estimating cross correlations from the problem of estimating mean returns and volatilities. In other words we will assume that the investor has a perfect forecast of $m_i$ and $\sigma_i$ and all her uncertainty is in the estimation of the cross correlation matrix. 

In order to quantify and compare the goodness of different filtering methods we make use of three measures. The first quantity measuring the reliability of the portfolio is obtained by comparing, for a given value of expected return, the risk $\sigma_p$ predicted by using the past correlation matrix with the realized risk $\hat\sigma_p$ of Eq.~\ref{realizedrisk}. A portfolio is more reliable when
\begin{equation}
{\cal R} \equiv \frac{|\hat\sigma_p -\sigma_p|}{\sigma_p}
\end{equation}  
is small. We have also used different measures of the reliability, such as $|\hat\sigma_p -\sigma_p|$, obtaining similar results. 

The second quantity used to compare different methods is simply the realized risk $\hat\sigma_p$. Clearly a portfolio is less risky of another one when its realized risk is smaller. Note that in general a portfolio with a small risk is not necessarily better than a more risky portfolio. In fact if the uncertainty on the risk of the safe portfolio is very large, an investor could face large fluctuations and therefore a larger loss.

The third characteristic for evaluating portfolio optimization methods is the degree of reduction in the effective dimension of the portfolio. Dealing with a large portfolio can be very costly because of the transaction costs that the investor has to face any time she wants to rebalance the weights. Even if we do not consider here the problem of portfolio rebalancing and benchmarking, we wish to quantify the ``effective" number of stocks with a significant amount of money invested in. We quantify this number as
\begin{equation}
{\cal N}^{(eff)}=\frac{1}{\sum_{i=1}^Np_i^2}
\label{partratio}
\end{equation}
This quantity is equal to $1$ when all the wealth is invested in only one asset, whereas it is equal to $N$ when the wealth is divided equally among the $N$ assets, i.e. $p_i=1/N$. It may be worth noting that the quantity ${\cal N}^{(eff)}$ does not give the number of assets where a non vanishing amount of wealth is invested in. It simply gives a rough estimate of the number of assets that could effectively be used to build a smaller portfolio with risk-returns properties not too far from the original $N$ asset portfolio.   

In the next sections, the different filtering procedures considered in this paper are investigated by using the set of data of $1071$ stocks continuosly traded at New York Stock Exchange (NYSE) during the period 1988-1998. In this study, we consider daily returns.

\section{Random Matrix Theory Approach}\label{rmtapp}

Recently \cite{Laloux99,Plerou99} it has been shown that the RMT can be useful to investigate the properties of return correlation matrices of financial assets. The simpler random matrix is a matrix of given type and size whose entries consist of random numbers from some specified distribution \cite{metha90}. RMT was developed originally in nuclear physics and then applied to many different fields. In the context of asset portfolios RMT is useful because allows to compute the effect of statistical uncertainty in the estimation of the correlation matrix. 
Suppose that the $N$ assets are described by $N$ time series of length $T$ and that the returns are  independent Gaussian random variables with zero mean and variance $\sigma^2$. The correlation matrix of this set of variables in the limit $T\to \infty$ is simply the identity matrix. When $T$ is finite the correlation matrix will in general be different from the identity matrix. RMT allows to prove that in the limit $T,N \to \infty$, with a fixed ratio $Q=T/N \geq 1$, the eigenvalue spectral density of the covariance matrix is given by
\begin{equation}
\rho(\lambda)=\frac{T}{2\pi\sigma^2\lambda}\sqrt{(\lambda_{max}-\lambda)
(\lambda-\lambda_{min})}, 
\label{zerofactor}
\end{equation}
where $\lambda_{min}^{max}=\sigma^2 (1+1/Q\pm 2\sqrt{1/Q})$. 
The spectral density is different from zero in the interval $]\lambda_{min},\lambda_{max}[$. In the case of a correlation matrix $\sigma^2=1$.
The spectrum described by Eq.~(\ref{zerofactor}) is different from $N\delta(\lambda-1)$ which is expected by an identity correlation matrix. In other words RMT quantifies the role of the finiteness of the length of the time series on the spectral properties of the correlation matrix. 

RMT has been applied to correlation matrices of returns of financial assets \cite{Laloux99,Plerou99} and it has been shown that the spectrum of a typical portfolio can be divided in three classes of eigenvalues. The largest eigenvalue is totally incompatible with Eq.~(\ref{zerofactor}) and describes the common behavior of the stocks composing the portfolio. A fraction of the order of $5\%$ of the eigenvalues are also incompatible with the RMT because they fall outside the interval   $]\lambda_{min},\lambda_{max}[$. These eigenvalues probably describe economic information stored in the correlation matrix. The remaining large part of the eigenvalues is between  $\lambda_{min}$ and $\lambda_{max}$ and thus one cannot say whether any information is contained in the corresponding eigenspace. 

The fact that by using RMT it is possible, under certain assumptions, to identify the noisy part of the correlation matrix suggested several authors \cite{Laloux00,Rosenow02} to use RMT in the optimization of financial portfolios. Specifically the suggested method \cite{Rosenow02} is the following. 
\begin{figure}[ptb]
\begin{center}
\includegraphics[scale=0.31,angle=-90]{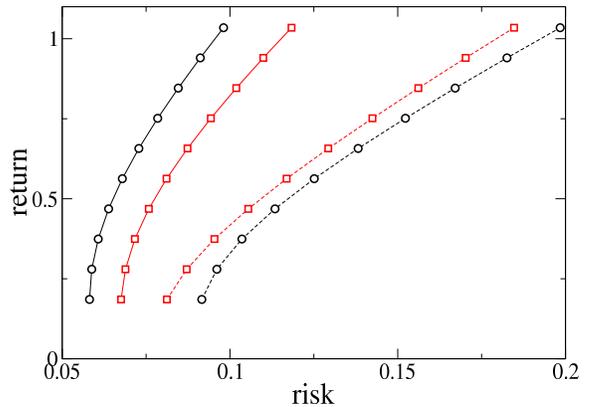}
\end{center}
\caption{The continuous  lines are the predicted risk and the dashed lines are the realized risk. The circles refer to the Markowitz portfolio optimization, whereas the squares are the predicted and realized risk curves obtained by filtering the correlation matrix with the Random Matrix Theory approach. We assume that the only uncertainty of the investor is on the correlation matrix. The dataset is composed by the $150$ most capitalized stocks at NYSE in the period $1989-1992$. The first two years are used for the estimation of the correlation matrix and the other two years are the investment period.}
\label{rmtfig}
\end{figure}

One computes the correlation matrix and finds the spectrum ranking the eigenvalues such that $\lambda_k<\lambda_{k+1}$. One then computes the variance of the part not explained by the highest eigenvalue as $\sigma^2=1-\lambda_1/N$ and uses this value in Eq.~\ref{zerofactor} to compute $\lambda_{min}$ and $\lambda_{max}$. One then constructs a filtered diagonal matrix obtained by setting to zero all the eigenvalues smaller than $\lambda_{max}$ and leaving unaltered the remaining ones. Finally one obtains the filtered correlation matrix by transforming the filtered diagonal matrix in the original basis. In order to obtain a meaningful correlation matrix we set to one the diagonal elements of the filtered correlation matrix. This matrix preserve only the information of the original correlation matrix that the RMT recognize as signal. It has been shown that the portfolio obtained by using the filtered correlation matrix has a smaller value of ${\cal R}$ than a portfolio with weights obtained with the Markowitz's procedure and by using the whole correlation matrix. As an example we show in Figure~\ref{rmtfig} the predicted and realized risk for a portfolio of the $150$ most capitalized stocks traded at NYSE in a period of $T=500$ trading days. In this and in the other figures the risk and return are computed on a yearly time horizon. We have estimated the correlation matrix in the two year period 1989-1990 and the realized risk is computed in the two year period 1991-1992. The figure shows the risk-return curve for the Markowitz portfolio and for a portfolio obtained by filtering the correlation matrix with the RMT method outlined above. We note that for all the value of the expected return $r_p$ the parameter ${\cal R}$ for the RMT portfolio is significantly smaller than for the Markowitz portfolio. For this portfolio the realized risk of the RMT portfolio is smaller than the realized risk of the Markowitz portfolio. This is in agreement with what, for example, Rosenow {\it et al.} \cite{Rosenow02} find for another set of data. We wish to point out that the behavior of Fig. 1 is rather common in different portfolios. However, we have found that for some portfolios the realized risk profile obtained with the RMT filtering is larger than the one obtained with the Markowitz approach.

\section{Clustering algorithms}\label{mst}

In this paper, we introduce a new portfolio optimization technique which is based on clustering algorithms. Clustering is a common practice in multivariate data analysis \cite{Mardia79}. The purpose of clustering analysis is to obtain a meaningful partition of a set of $N$ variables in groups according to their characteristics. For example in correlation based clustering algorithms (adopted here) the correlation coefficient between two time series is assumed as a measure of the similarity between the two time series. Correlation based clustering has been recently used to infer the hierarchical structure of a portfolio of stocks from its correlation coefficient matrix \cite{Mantegna99,Bonanno01,Bonanno03}. Correlation based clustering may be seen as a filtering procedure, i.e. a matrix transformation retaining a smaller number of distinct elements. After its application one usually retain a subset of the distinct elements composing the correlation coefficient matrix. For example, in the clustering algorithm of the single linkage \cite{Gower69} the number of distinct elements present in the filtered matrix is $n-1$ whereas the number of distinct elements present in the original matrix is $n (n-1)/2$. The selection of these $n-1$ elements is done according to some widespread algorithm \cite{Papadimitriou82}. A possible conceptual description of the algorithm is the following. Let us assume that a similarity measure $S$ between pairs of elements is defined, e.g. the correlation coefficient between pairs of elements of the system. An ordered list $S_{ord}$ of pair of elements can be constructed  by arranging them in a descending order accordingly with the value of the similarity $s_{ij}$ between element $i$  and element $j$. 
Different elements are iteratively included in clusters starting from the first two elements of the similatity measure ordered list. At each step, when two elements or one element and a cluster or two clusters $p$ and $q$ merge in a wider single cluster $t$, the similarity or distance between the new cluster $t$ and cluster $r$ is determined as follows: if $s_{ij}$ is a correlation-like measure 
\begin{equation}
s_{tr}=\max \{s_{pr},s_{qr}\}
\end{equation}
indicating that the similarity between any element of cluster $t$ and any element of cluster $r$ is the similarity between the two most similar entities in clusters $t$ and $r$. Conversely, if $s_{ij}$ is a distance-like measure 
\begin{equation}
s_{tr}=\min \{s_{pr},s_{qr}\}.
\end{equation}

By applying iterarively this procedure $n-1$ of the $n(n-1)/2$ distinct elements of the correlation coefficient matrix are selected. When a distance-like measure is used as, for example, $d_{ij}=\sqrt{2(1-\rho_{ij})}$ \cite{Gower66}, the distance matrix obtained by applying the single linkage procedure is an ultrametric matrix comprising the $n-1$ distinct selected elements. Ultrametric distances $d_{ij}^<$ are distances satisfying a inequality $d^<_{ac} \le  \max \{d^<_{ab},d^<_{bc}\}$ stronger than the customary triangular inequality $d_{ac} \le  d_{ab} + d_{bc}$ \cite{Rammal86}. In particular, the single linkage clustering procedure has associated an ultrametric correlation coefficient matrix which is the subdominant ultrametric matrix of the original correlation coefficient matrix. For a didactic description of the method used to obtain the ultrametric matrix one can consult Ref. \cite{MS00}.  

In Ref. \cite{TumminelloManuscript} it is proved that the ultrametric correlation matrix obtained by the single linkage clustering procedure of the correlation coefficient matrix is always positive definite when all the elements of the obtained ultrametric correlation matrix are positive. This condition is rather common in financial data of stock portfolio and it has always been observed for all the investigations we have performed so far. The effectiveness of the single linkage clustering procedure in pointing out the hierarchical structure of the investigated portfolio has been shown by several studies \cite{Mantegna99,Bonanno00,Bonanno01,Kullmann02,Onnela02,Micciche03,Bonanno03,Bonanno04,Dimatteo04}. However, the single linkage is just one possible correlation based filtering methods. Other methods have also been applied to financial portfolios \cite{Kullmann00,Giada01,Maslov01,Bernaschi02,Kullmann02,Onnela02,Mendes03,Maskawa03,Basalto05,Tumminello05}. Each method puts a specific emphasis on some aspects of the original matrix and is usually able to point out a series of aspects that might not be elucidated by different filtering procedure. The choice of the filtering method must therefore be guided by the specific goals that one pursues. In the present study, we decide to consider the average linkage procedure in addition to the single linkage procedure. The average linkage is another widespread clustering algorithm \cite{Anderberg73}. The difference with the single linkage algorithm is that the similarity measure between an element and the closest cluster is given by the mean similarity measure between the considered element and each element of the closest cluster. 
In other words, if $s_{ij}$ is a similarity-like measure, at each stage one obtains $s_{tr}$ between clusters $t$ and $r$ defined as above, as the average distance between all pairs of links of the elements belonging to the two clusters. For a detailed discussion of the average linkage cluster algorithm see, for example Ref. \cite{Anderberg73}. Also in the case of the average linkage the filtered correlation coefficient matrix is a ultrametric distance. In Ref. \cite{TumminelloManuscript} it is proved that the ultrametric correlation coefficient matrix associated with the average linkage clustering procedure is positive definite under the same general conditions valid for the case of the single linkage. However, this property is not generic to all clustering procedures. We have verified that it does not apply for the cases of the complete linkage and for the Ward clustering method.

\section{Portfolio optimization with clustering algorithms}\label{method}

The portfolio optimization method we propose here is based on the use of the ultrametric matrix associated to a given clustering method as a meaningful and robust filtration of the original correlation matrix. In other words we construct the portfolio by solving the Markowitz optimization problem by using the ultrametric matrix rather than the original correlation matrix or the RMT filtered matrix. The reasons for this choice are (i) it is known that clustering algorithms are able to filter the relevant information in a multivariate set of data (ii) other studies indicate that clustering algorithms are quite robust with respect to measurement noise due to the finiteness of sample size. This is particularly true for set of variables hierarchically organized \cite{TumminelloManuscript}.  

From the filtered (ultrametric) correlation matrix we build the portfolio by using the Markowitz result (Eq.~\ref{markowitz}) and for each value of the portfolio return $r_p$ we find the predicted risk $\sigma_p$. We note that in order to consider the ultrametric matrix as a meaningful correlation matrix it is important that the matrix is positive definite (or semidefinite).  We have performed a very large number of portfolio optimization using real data and we have not found a single case in which the ultrametric matrix is not positive definite.  We used the weights obtained from the optimization to compute the realized risk by using Eq.~\ref{realizedrisk} where $\hat\sigma_{ij}$ is the {\it original} covariance matrix. In other words we use the filtered matrix only for obtaining the weights $p_i$, whereas the realized risk is clearly determined by the whole correlation matrix.

\subsection{Average linkage}

We consider first portfolio built by using correlation matrices filtered with the average linkage cluster algorithm.

\subsubsection{Reliability}
\begin{figure}[ptb]
\begin{center}
\includegraphics[scale=0.31,angle=-90]{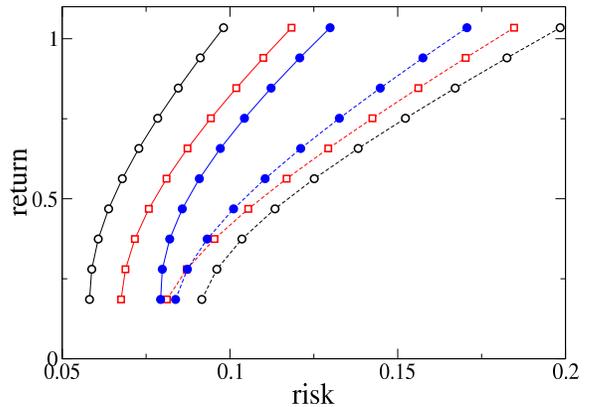}
\end{center}
\caption{The continuous  lines are the predicted risk and the dashed lines are the realized risk. The filled blue circles refer to the portfolio obtained with the average linkage method. The empty circles refer to the Markowitz portfolio optimization, whereas the squares are the predicted and realized risk curves obtained by filtering the correlation matrix with the Random Matrix Theory approach (same data as in Fig.~\ref{rmtfig}). We assume that the only uncertainty of the investor is on the correlation matrix. The dataset is composed by the $150$ most capitalized stocks at NYSE in the period $1989-1992$. The first two years are used for estimate the correlation matrix and the other two years are the investment period.}
\label{avgfig}
\end{figure}

Figure~\ref{avgfig} shows the predicted and realized risk for the portfolio obtained with the average linkage considering the same set of stocks and the same time period as in Fig.~\ref{rmtfig}. The distance between predicted and realized risk for the portfolio obtained with average linkage is significantly smaller than the distance for the portfolio obtained with the RMT. This result indicates clearly that the use of clustering methods to build financial portfolio is able to provide portfolios more reliable (in terms of the error in the forecasted risk) than the ones obtained with RMT and with Markowitz optimization.  We also note that for this set of data the realized risk of the portfolio obtained with the clustering method is almost always smaller than the realized risk of the RMT portfolio.

\begin{figure}[ptb]
\begin{center}
\includegraphics[scale=0.3]{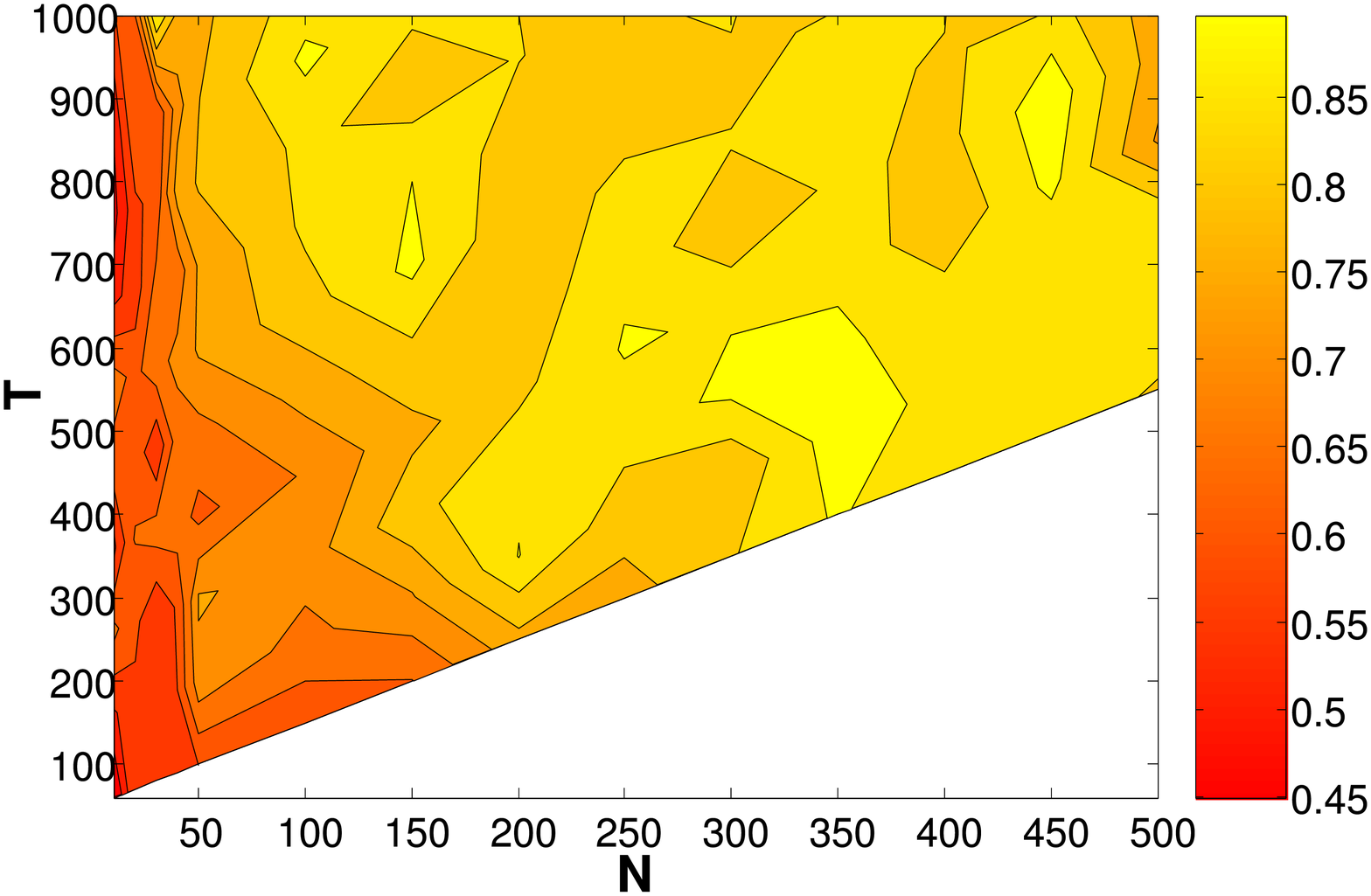}
\end{center}
\caption{Density plot showing the percentage of success of the average linkage portfolio optimization technique over the Random Matrix Theory approach as a function of the number of asset $N$ and the investment period $T$. The white area corresponding to cases where $T<N+50$ is not considered in our investigation.}
\label{avgrmtdensityfig}
\end{figure}

In order to verify the robustness of these results we have performed an extensive bootstrap experiment. We have considered many different values of the portfolio size $N$ and of the investment horizon $T$ and for each couple $(N,T)$ we have randomly sampled $50$ portfolio composed by $N$ stocks and we have selected randomly $50$ initial times $t_0$. For each portfolio we have considered $10$ values of the expected portfolio return $r_p$. Specifically we have taken ten equispaced values of $r_p$ between the  value of $r_p$ associated with the absolute minimum risk and the highest expected return among the $N$ stocks of the portfolio. For each expected return and for each portfolio we have computed the parameter ${\cal R}$  and we have counted the fraction of times that ${\cal R}_{av.link}<{\cal R}_{RMT}$, i.e. the percentage of cases in which the portfolio obtained with average linkage is more reliable than the portfolio obtained with Random Matrix Theory. The result of this analysis is shown in Figure~\ref{avgrmtdensityfig}.   The figure indicates that the average linkage portfolio outperforms the RMT portfolio almost for any value of $N$ and $T$. For $N\simeq 350$ and $T\simeq 500$ the average linkage portfolio is more reliable than the RMT portfolio more than $85\%$ of the times. The reliability of average linkage portfolio is higher when the number of stocks is large, For small size ($N<50$) portfolios the two methods are statistically equally reliable. 

\subsubsection{Riskiness}
We now compare the realized risk of the portfolios obtained with the two methods, i.e. RMT and average linkage. The realized risk is a measure of the riskiness of the portfolio. 
We observe that small size portfolio tends to be less risky when obtained with the average linkage, whereas as $N$ increases the RMT portfolios become less risky. The boundary between the two regions is approximately for $N\sim75$. 
By comparing this result with figure ~\ref{avgrmtdensityfig} we see that when the average linkage is more reliable it is also riskier and vice-versa. There is a small region around $N\simeq50$ where it is possible to build portfolio with average linkage which are reliable and not too risky. 

It is important to stress that, as for Figure~\ref{avgrmtdensityfig}, the above result on riskiness is obtained by putting together all the values of portfolio expected return $r_p$. On the other hand we find that the riskiness of average linkage portfolio compared to RMT portfolio strongly depends on $r_p$ especially for large portfolios. Specifically when we consider large portfolios ($50<N<500$) we find that for small $r_p$ only in $\sim 25\%$  of the cases the average linkage portfolio is less risky than RMT portfolio. When $r_p$ is large this fraction is of the order of $\sim 45\%$. In other words for portfolios with large $r_p$ the average linkage portfolios is approximately as risky as the RMT portfolios.

\subsubsection{Effective size}

\begin{figure}[ptb]
\begin{center}
\includegraphics[scale=0.33,angle=-90]{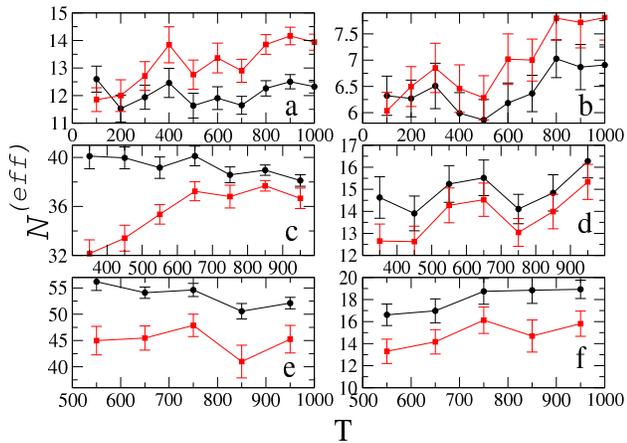}
\end{center}
\caption{Effective size ${\cal N}^{(eff)}$ of the porfolio as defined in Eq.~\ref{partratio} as a function of investment horizon $T$. The black circles refer to RMT portfolio and the red squares to average linkage portfolio. The portfolio size is $N=50$ (panels (a) and (b)), $N=300$ (panels (c) and (d)) and $N=500$ (panels (e) and (f)). The left panels (a,c, and e) refer to the minimum value of portfolio expected return $r_p$ and the right panels (b,d, and f) refer to an intermediate value of $r_p$. Every point is the average over $50$ realizations obtained by bootstrapping and the error bars are standard errors.}
\label{pratioavg}
\end{figure}

Finally we consider the effective size ${\cal N}^{(eff)}$ of the portfolio as quantified by Eq.~\ref{partratio}. We consider three portfolio sizes, i.e. $N=50, 300$, and $500$, and we select two values of the portfolio expected return $r_p$, i.e. the minimum value (corresponding to the minimum risk) and an intermediate value between the minimum and the maximum. Figure~\ref{pratioavg} shows ${\cal N}^{(eff)}$ as a function of the investment horizon $T$. Similar results are observed for high values of expected return, but in this case the dimensionality of the portfolio becomes smaller and smaller and the wealth is more and more concentrated in the asset with highest return.
We note that for small portfolio ($N=50$) the effective size of RMT portfolios is slightly smaller than the effective size for average linkage portfolios. On the other hand for larger portfolios the effective size of average linkage portfolios is significantly smaller than the effective size of RMT portfolios. This result shows that portfolios built with average linkage have a smaller effective dimensionality, i.e. the maintenance cost of these portfolios is smaller than for the one of RMT portfolios.

\subsection{Single linkage}

\begin{figure}[ptb]
\begin{center}
\includegraphics[scale=0.31,angle=-90]{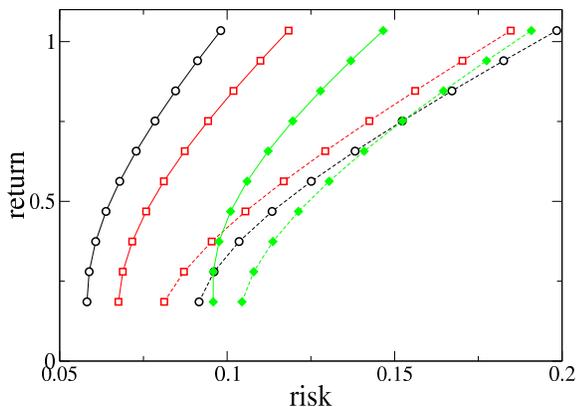}
\end{center}
\caption{The continuous  lines are the predicted risk and the dashed lines are the realized risk. The filled green diamonds refer to the portfolio obtained with the single linkage method. The empty circles refer to the Markowitz portfolio optimization, whereas the squares are the predicted and realized risk curves obtained by filtering the correlation matrix with the Random Matrix Theory approach (same data as in fig.~\ref{rmtfig}). We assume that the only uncertainty of the investor is on the correlation matrix. The dataset is composed by the $150$ most capitalized stocks at NYSE in the period $1989-1992$. The first two years are used for estimate the correlation matrix and the other two years are the investment period.}
\label{mstfig}
\end{figure}

\subsubsection{Reliability}

We performed the same analysis by using a different clustering algorithm, specifically the single linkage cluster analysis. By using the same data as in Figures~\ref{rmtfig} and \ref{avgfig} we compute the curves for predicted and realized risk for a portfolio built by using the ultrametric matrix associated with the single linkage algorithm. The result is in Fig.~\ref{mstfig}. Also in this case the predicted and realized risk for single linkage portfolio are significantly closer than the corresponding quantities for Markowitz and for RMT portfolios. In this case it is more evident that the realized risk of the single linkage portfolio is larger than the other two realized risks. Thus the single linkage portfolio is riskier but more reliable when compared with RMT portfolio.

\begin{figure}[ptb]
\begin{center}
\includegraphics[scale=0.3]{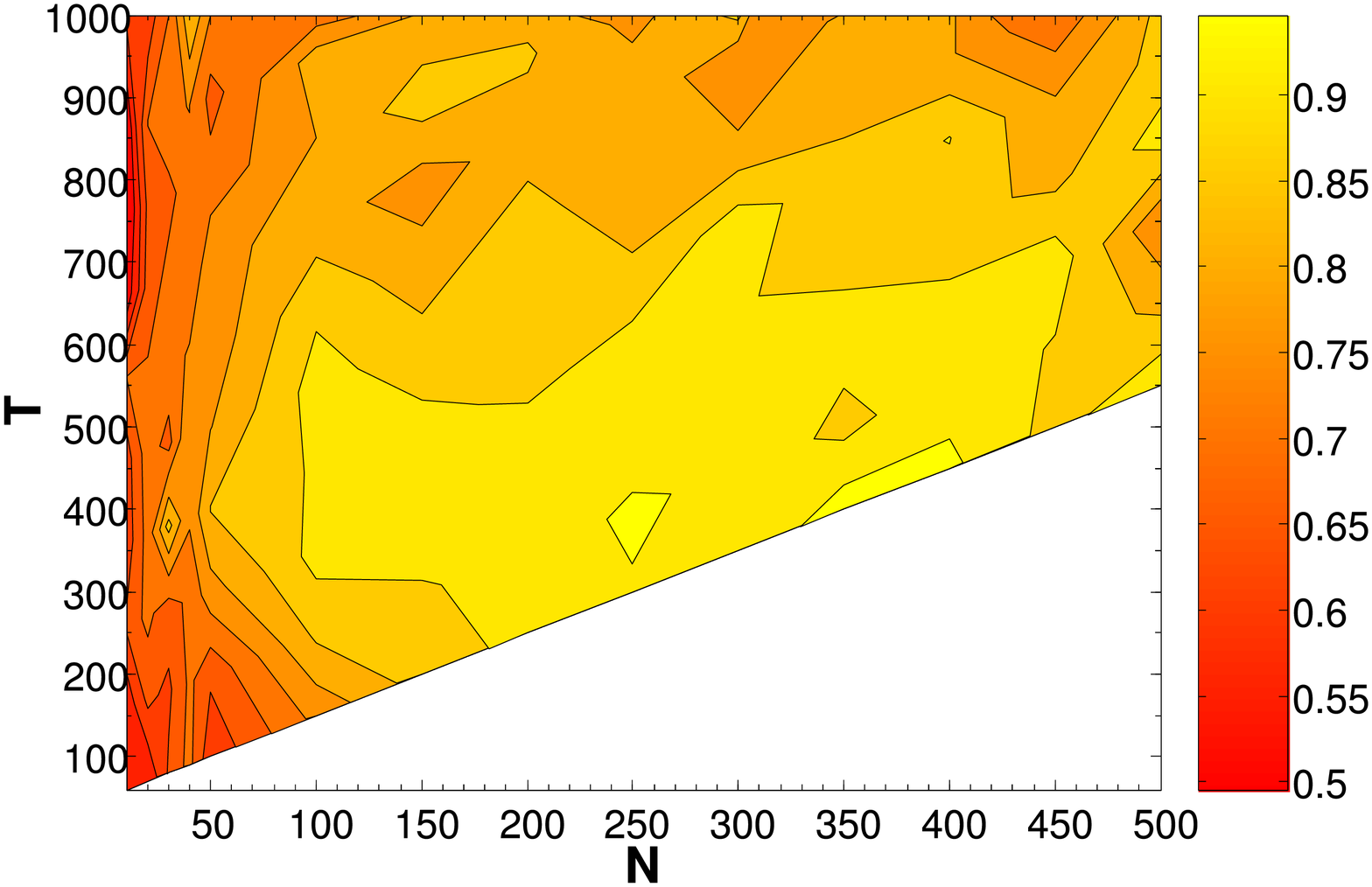}
\end{center}
\caption{Density plot showing the percentage of success of the single linkage portfolio optimization technique over the Random Matrix Theory approach as a function of the number of asset $N$ and the investment period $T$. The white area corresponding to cases where $T<N+50$ is not considered in our investigation.}
\label{mstdensityfig}
\end{figure}

Also for the single linkage method we perform a bootstrap analysis similar to the one described above for the average linkage method in order to compare the reliability of single linkage method as compared to RMT method. The result is the density plot shown in Fig.~\ref{mstdensityfig}.
We find that the single linkage method is able to provide more reliable portfolios in wide ranges of the parameters $N$ and $T$. This is more and more evident for portfolios with $T\simeq N$, i.e. portfolios for which the investment horizon (in trading days) is comparable with the portfolio size $N$. It is interesting to note that for these portfolio the effect of the measurement noise (``noise dressing") is particularly high. 

\subsubsection{Riskiness}

The analysis of the riskiness, i.e. the realized risk, of single linkage portfolios shows that these portfolio are systematically riskier than RMT portfolios. This is also seen in the example shown in figure~\ref{mstfig}. Only for very small size ($N<15$) the single linkage portfolios are less risky than RMT portfolios. As for the average linkage portfolio this effect strongly depends on the portfolio expected return. When we consider large portfolios ($50<N<500$) we find that for small $r_p$ only in $\sim 0.3\%$  of the cases the single linkage portfolio is less risky than RMT portfolio. When $r_p$ is large this fraction is of the order of $\sim 10\%$. In any case these figures indicate that single linkage portfolios are risky, even if they can be quite reliable.

\subsubsection{Effective size}

\begin{figure}[ptb]
\begin{center}
\includegraphics[scale=0.33,angle=-90]{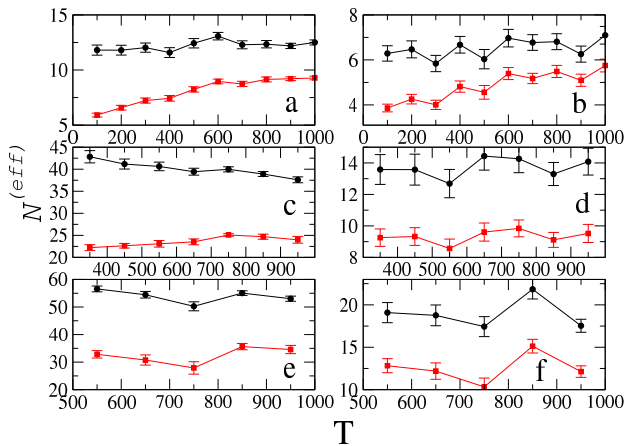}
\end{center}
\caption{Effective size ${\cal N}^{(eff)}$ of the porfolio as defined in Eq.~\ref{partratio} as a function of investment horizon $T$. The black circles refer to RMT portfolio and the red squares to single linkage portfolio. The portfolio size is $N=50$ (panels (a) and (b)), $N=300$ (panels (c) and (d)) and $N=500$ (panels (e) and (f)). The left panels (a,c, and e) refer to the minimum value of portfolio expected return $r_p$ and the right panels (b,d, and f) refer to an intermediate value of $r_p$. Every point is the average over $50$ realizations obtained by bootstrapping and the error bars are standard errors.}
\label{pratiomst}
\end{figure}

The effective size ${\cal N}^{(eff)}$ of the portfolio as quantified by Eq.~\ref{partratio} for the single linkage portfolio shows interesting properties. Figure~\ref{pratiomst} shows the effective size for different portfolio conditions and should be compared with figure~\ref{pratioavg}. We see that for any value of the portfolio size $N$ the effective size of the single linkage portfolio is significantly smaller than the one of the RMT portfolio. This effect is observed also for small size portfolios, in contrast with what we observe for average linkage portfolios. Even more important, for large portfolios (where the size reduction is an important issue) the single linkage portfolios have an effective size which is roughly half the effective size of RMT portfolio. This result suggest that single linkage portfolios could be used to detect a small subset of stocks which is representative of the whole portfolio, and thus to replace the original portfolio with another one of significantly smaller size with reduced maintenance costs. This possibility will be explored in a future work. 

\section{Conclusions}\label{conclusions}

In this paper we have performed portfolio optimization by using 
filtered correlation coefficient matrices. These matrices have been
obtained by applying different filtering methods to the original 
correlation coefficient matrix. 
We have proposed two filtering methods based on the average linkage
and single linkage clustering procedures. The optimal portfolios
obtained with these two new methods have been compared with the 
one based on RMT recently proposed in Refs \cite{Laloux00,Rosenow02}.

A large set of simulations
has shown that clustering methods are outperforming RMT filtering
when we consider the reliability of the estimation of the realized
portfolio with respect to the predicted one for portfolios with a 
number of assets $ \approx 50 < N < \approx 500$.
Hence, for relatively large portfolios the clustering filtering methods
provide a more reliable estimation of the predicted risk-return profile
both with respect to the Markowitz basic estimation and
with respect to the determination of the correlation 
coefficient done with the RMT filtering

The portfolios obtained with the average linkage shows a predicted and realized risk return profile which is often inside the corresponding profiles obtained both with the Markowitz basic estimation and
after the RMT filtering. In the case of the single linkage
clustering method the risk-return profile shows risk levels which 
are systematically higher than the ones obtained both with the 
Markowitz basic estimation and after the RMT filtering. 
Therefore with respect to the aspect of the level of risk
associated to the selected portfolios the most successful
methods are the average linkage and the RMT filtering.

Another aspect investigated in our study refers to the 
composition of the portfolios selected. We have quantified 
the degree of homogeneity of the distribution of the
wealth across the stocks of the portfolio through what we have
called the "effective size" of the portfolio. A small number
of this parameter indicates an uneven distribution of the 
portfolio wealth suggesting that during portfolio
re-balancing only a subset of stocks will be significantly 
involved. The investigation of the "effective size" of the
portfolio has shown that the average linkage and the RMT are
characterized by not too different values of the "effective size".
In fact for small portfolios (e.g. $N=50$) the RMT has for most
values of $T$ a smaller value of the "effective size" whereas the
pattern is reversed for medium ($N=300$) and large 
portfolios ($N=500$) both for the minimum and intermediate
value of $r_p$. The pattern is clearly different for the 
case of the single linkage filtering. In this case the 
"effective size" is always significantly less than the 
one observed in the cases of RMT filtering.

The above discussion shows that the different filtering 
procedures provide different portfolio optimization results
that are characterized by specific strengths or
weaknesses. In other words, for each value of $N$ and 
$T$, the most useful filtered correlation coefficient matrix
can be different depending on the strongest 
constraint the investor has among the risk
level of the portfolio, the reliability of the estimation
and the portfolio "effective size".
 We believe that the two clustering methods 
we have proposed here and the RMT are not exhaustive 
with respect to all potential aspects of portfolio
optimization and, probably, other filtering methods 
could also provide very interesting results in specific
regimes of the different control parameters.

The different results of the different filtering methods
raise the scientific question of which is the reason 
for the difference between the various filtering
procedures. A precise quantification of the information
retained by the different filtered matrices would be
very useful. This goal is left for future research.

\acknowledgments Authors acknowledge support from the research project MIUR 449/97 ``High frequency dynamics in financial markets" . F.L. and R.N.M  acknowledge support also from the research project MIUR-FIRB RBNE01CW3M ``Cellular Self-Organizing nets and chaotic nonlinear dynamics to model and control complex system'' and from the European Union STREP project n. 012911 ``Human behavior through dynamics of complex social networks: an interdisciplinary approach.''.

\end{document}